\documentclass[12pt]{article}

\catcode`\@=11

\global\arraycolsep=2pt
\usepackage{amsbsy,amssymb,latexsym,amsfonts,amsmath}
\usepackage{graphicx,color}

\begin{document}
\begin{flushright}
\parbox{4.2cm}
{IPMU 14-0011}
\end{flushright}

\vspace*{0.7cm}

\begin{center}
{\Large $a-c$ test of holography vs quantum renormalization group}
\vspace*{1.5cm}\\
{Yu Nakayama}
\end{center}
\vspace*{1.0cm}
\begin{center}
{\it Kavli Institute for the Physics and Mathematics of the Universe (WPI),  \\ Todai Institutes for Advanced Study,
University of Tokyo, \\ 
5-1-5 Kashiwanoha, Kashiwa, Chiba 277-8583, Japan}
\vspace{3.8cm}
\end{center}

\begin{abstract}
We show that a ``constructive derivation" of the AdS/CFT correspondence based on the quantum local renormalization group in large $N$ quantum field theories consistently provides the $a-c$ holographic Weyl anomaly in $d=4$ at the curvature squared order in the bulk action. The consistency of the construction further predicts the form of the metric beta function.
\end{abstract}

\thispagestyle{empty} 

\setcounter{page}{0}

\newpage

S.~S.~ Lee recently proposed a constructive way to obtain the $d+1$ dimensional bulk action from the quantum local renormalization group in large $N$ quantum field theories in $d$ dimension to aim at the derivation of the AdS/CFT correspondence \cite{Lee:2012xba}\cite{Lee:2013xba} (see also the related idea in \cite{Verlinde:1999xm}\cite{Kiritsis:2012ta}). In this short article, we would like to give a modest but concrete consistency check of his proposal by comparing unambiguously computable quantities both in field theory and gravity.

Let us consider a (hypothetical) large $N$ quantum field theory in $d=4$ dimension, where the single-trace energy-momentum tensor and its multi-trace cousins are the only operators with finite scaling dimension. We also assume that the theory is ``strongly coupled" in the sense that the higher derivative terms beyond the background curvature squared  are suppressed in the local renormalization group. The existence of the other operators would not change the following story as long as they can be consistently ``decoupled" within the computations of the energy-momentum tensor correlation functions (e.g. strongly coupled $\mathcal{N}=4$ super Yang-Mills theory).

By using the recipe proposed by Lee \cite{Lee:2013xba}, we can formally rewrite the Schwinger functional, which is identified as the GKP/W partition function,  in terms of the bulk $d+1 = 5$ dimensional path-integral 
\begin{align}
e^{-W[g_{\mu\nu}]} = \int \mathcal{D}X e^{-\int d^4x \sqrt{g} L(X;g_{\mu\nu})} = \int \frac{\mathcal{D} g_{\mu\nu}\mathcal{D}\pi^{\mu\nu} \mathcal{D}n \mathcal{D}n^\mu}{\mathrm{Diff}} e^{-N^2S_B} \ , \label{bulkpath}
\end{align}
where the bulk action in the Hamiltonian formulation takes the form
\begin{align}
S_B = \int d^4x dz \left( \pi^{\mu\nu} \partial_z g_{\mu\nu} - n^\mu H_\mu - n H \right)
\end{align}
with the $4+1$ dimensional metric $ds^2 = G_{MN} dx^M dx^N = n^2 dz^2 + g_{\mu\nu} (dx^\mu + n^\mu dz)(dx^\nu + n^\nu dz)$. The GKP-W boundary condition naturally follows from the construction of  \cite{Lee:2013xba} in the asymptotic AdS case.

The Hamiltonian density $H$ is determined by the renormalization group properties of the dual field theory 
\begin{align}
H = \sqrt{g}\Lambda[g_{\mu\nu}] - \beta_{\mu\nu}[g_{\mu\nu}] \pi^{\mu\nu} - \mathcal{G}_{\mu\nu;\rho\sigma} [g_{\mu\nu}]  \pi^{\mu\nu} \pi^{\rho\sigma} \ . \label{action}
\end{align}
Here $\Lambda[g_{\mu\nu}]$ is determined from the renormalization of the ``cosmological constant" in the dual field theory.
$\beta_{\mu\nu}[g_{\mu\nu}]  $ is the beta function for the single trace energy-momentum tensor $:T_{\mu\nu}:$, and  $\mathcal{G}_{\mu\nu;\rho\sigma} [g_{\mu\nu}] $ is the beta function for the double trace energy-momentum tensor  $:T_{\mu\nu} T_{\rho\sigma}:$ of the dual field theory respectively. Our normalization is such that the single trace operator is $O(1)$.

Before we proceed, let us emphasize two important properties of this bulk action $S_B$ and its path integral in \eqref{bulkpath}.
First of all, we note that the Schwinger functional by construction does not depend on the choice of the local renormalization scale $n(x^\mu,z ) dz = d\Lambda(x^\mu)$. The arbitrariness of the local renormalization scale is assured from the local renormalizability, and it is manifested as the local Callan-Symanzik equation (with Weyl anomaly included). 
In the bulk language, this is interpreted as the Hamiltonian constraint  $H=0$ from the variation of the Lapse function $n$ because the Hamiltonian constraint guarantees the local reparametrization of the radial direction, which we identify as the local renormalization group transformation. Similarly, after varying the action with the shift vector $n^\mu$, $H_\mu = -2D^\nu \pi_{\mu\nu} = 0 $ gives the momentum constraint, whose origin is the condition that the $d=4$ diffeomorphism invariance is preserved under the local renormalization group.

The second point to note is the linear term $\beta_{\mu\nu} \pi^{\mu\nu}$ in $H$. We are eventually interested in the Lagrangian formulation of the bulk action by integrating out the ``canonical momentum" $\pi^{\mu\nu}$, but the appearance of the linear term generically prevents us from obtaining the explicitly $d+1$ diffeomorphism invariant Lagrangian because it could result in the first order derivative terms in the $z$ direction. As pointed out in \cite{Lee:2012xba}, if $\beta_{\mu\nu}$ is generated by a gradient flow, the resulting first order derivative term $G^{\mu\nu;\rho\sigma} \beta_{\mu\nu} \partial_z g_{\rho\sigma}$ is a total derivative, where 
$G^{\mu\nu;\rho\sigma}$ is the inverse of the double trace beta function $\mathcal{G}_{\mu\nu;\rho\sigma}$, and can be removed as a boundary term. We will come back to this point toward the end of this article. By assuming the gradient condition, integrating out $\pi_{\mu\nu}$ will also generate the potential term $\beta_{\mu\nu} G^{\mu\nu;\rho\sigma} \beta_{\rho\sigma}$.

So far, our argument is quite general, but now we would like to perform the consistency check of the constructed bulk action at $O(R^2)$ in curvature expansion to compare it with the holographic prediction. As we will explain, our main focus will be the holography $a-c$ Weyl anomaly and the associated $R_{IJKL}^2$ term in the bulk action in the Lagrangian formulation, so we pay particular attention to the origin of this term.

We are going to keep $O(R^2)$ term in $\Lambda[g_{\mu\nu}]$ and $O(R)$ term in $\beta_{\mu\nu}[g_{\mu\nu}] $ and $\mathcal{G}_{\mu\nu;\rho\sigma}[g_{\mu\nu}] $ because the comparison with the holographic Weyl anomaly we would like to perform assumes that the bulk action is $O(R^2)$ in the Lagrangian formulation after integrating out $\pi^{\mu\nu}$. Explicitly, we have
\begin{align}
\Lambda[g_{\mu\nu}] &= \Lambda_0 + R + \alpha_1 R^2 + \alpha_2 R_{\mu\nu}^2 + \alpha_3 R_{\mu\nu\rho\sigma}^2  + \alpha_4 \Box R \cr
\beta_{\mu\nu} & = 2 g_{\mu\nu} + \beta_1 R g_{\mu\nu} + \beta_2 R_{\mu\nu} \cr
 \mathcal{G}_{\mu\nu;\rho\sigma}& = \frac{1}{\sqrt{g}} (g_{\mu\rho} g_{\nu\sigma} - \lambda g_{\mu\nu} g_{\rho\sigma}) + O(R) \ . \label{beta}
\end{align}

The curvature expansion here is standard, but we would like to make a comment on the justification of the truncation for each terms. In order to perform a meaningful consistency check, we need a strict control of the higher curvature terms in the bulk action because the contribution to the holographic anomaly from the higher curvature terms are generically not suppressed at all. We stress this is not the restriction of the general construction of the local renormalization group, but it is due to our lack of our understanding in the bulk computation of the holographic Weyl anomaly.  

The truncation in \eqref{beta} is sufficient in the sense that after integrating out $\pi_{\mu\nu}$, the resulting action in the Lagrangian formulation only contains terms up to $O(R^2)$. In particular, the crucial point we will employ in the following  is that $R_{\mu\nu\rho\sigma}^2$ contribution to the bulk action only comes from $\Lambda[g_{\mu\nu}]$ within the order we are interested in.

To see the necessity, we only consider the possible higher terms in \eqref{beta} that would contribute to $O(R^2)$ terms after integrating out $\pi_{\mu\nu}$. The $O(R^3)$ terms in $\Lambda(g_{\mu\nu})$ are thus irrelevant. $O(R^2)$ in $\beta_{\mu\nu}$ could contribute to the $O(R^2)$ action from $\beta_{\mu\nu} G^{\mu\nu;\rho\sigma} \beta_{\rho\sigma}$, but we note that this would not contribute to the $R_{\mu\nu\rho\sigma}^2$ term. In addition, we generically obtain $O(R^3)$ terms that may or may not be cancelled from $\Lambda[g_{\mu\nu}]$. Finally, $O(R)$ term of $\mathcal{G}_{\mu\nu;\rho\sigma}$ suppressed in \eqref{beta} should not contribute to the  $R_{\mu\nu\rho\sigma}^2$ term. If we considered $R^2$ order in (the inverse of) $\mathcal{G}_{\mu\nu;\rho\sigma}$, we could encounter the additional $R_{\mu\nu\rho\sigma}^2$ term, which is not originated from $\Lambda[g_{\mu\nu}]$. However, it should necessarily accompany higher derivative curvature terms from the kinetic term $ \partial_z g_{\mu\nu} G^{\mu\nu;\rho\sigma} \partial_z g_{\rho\sigma}$  of $O(R^3)$ that {\it cannot} be cancelled from the other terms, 
and it would be beyond our scope of $O(R^2)$ test of the holographic Weyl anomaly. We therefore conclude that the truncation \eqref{beta} is necessary and sufficient for the $O(R^2)$ check of holography.\footnote{We should note that it is quite non-trivial how this truncation actually happens in a given quantum field theory, which is related to the question of the meaning of ``strongly coupled" in the AdS/CFT correspondence.}

 It is also important to realize that the field redefinition of the type $g_{\mu\nu} \to g_{\mu\nu} + \zeta_1 R_{\mu\nu} + \zeta_2 R g_{\mu\nu}$ associated with the renormalization group scheme ambiguity cannot affect the $R_{\mu\nu\rho\sigma}^2$ term.
There should be additional consistency conditions so that the bulk action is $d+1$ dimensional diffeomorphism invariant. This gives various relations and fixes some parameters such as $\lambda = \frac{1}{3}$ in the last line of \eqref{beta}.

Most of these coefficients in \eqref{beta} cannot be computed in the power-counting renormalization scheme of the dual field theory (while it should be done in Wilsonian framework), but the exception is $\alpha_1$, $\alpha_2$, $\alpha_3$ and $\alpha_4$. These dimensionless terms are fixed by the Weyl anomaly up to local counterterms so that the two particular combinations (so-called $a$ and $c$) do not depend on the renormalization scheme and they are universal (see e.g. \cite{os} \cite{Nakayama:2013is} and reference therein for the local renormalization group within the power-counting renormalization scheme). Let us assume that the dual field theory (at the fixed point) is a conformal field theory. The conformal fixed point is characterized by the central charges $a$ and $c$, and the Weyl anomaly gives the renormalization of the cosmological constant term in the Schwinger functional as
\begin{align}
\Lambda_{\mathrm{anomaly}}[g_{\mu\nu}] &= c\mathrm{Weyl}^2 - a\mathrm{Euler} \cr
 & = \left(\frac{c}{3} - a\right) R^2 + (-2c + 4a) R_{\mu\nu}^2 + (c-a)R_{\mu\nu\rho\sigma}^2 \ , \label{anomaly}
\end{align}
which determines a part of the bulk action \eqref{action} according to the quantum local renormalization group recipe. We have used the local counterterm to get rid of $\Box R$ term, but it does not affect the following argument. Since the part of the bulk action is fixed by the Weyl anomaly, we would like to check if the so-constructed bulk action gives back the holographic Weyl anomaly in a consistent manner.

Let us first recall that the holographic Weyl anomaly predicts that the field theory dual to Einstein gravity must have $a=c$. At the second order in the bulk curvature, the most general possibility of the bulk action in the Lagrangian formulation is
\begin{align}
S_2 = \int d^4x dz \sqrt{G} (\lambda_1 (R^{(5)})^2 + \lambda_2 (R_{MN}^{(5)})^2 + \lambda_3 (R_{IJKL}^{(5)})^2) \ . \label{bulks}
\end{align}
The computation of the holographic Weyl anomaly with these curvature squared terms are done in \cite{Nojiri:1999mh}\cite{Blau:1999vz}, and the salient feature of their result is that the difference $a-c$ in the holographic Weyl anomaly is only induced by $\lambda_3 (R_{IJKL}^{(5)})^2$ term as $a-c \sim N^2 \lambda_3 $ and does not depend on $\lambda_1$ and $\lambda_2$ except through the change of the overall AdS radius. However, this is exactly what is proposed by the quantum local renormalization group construction because $(R_{IJKL}^{(5)})^2$ term in the bulk action is in one to one correspondence with $R_{\mu\nu\rho\sigma}^2$ term in $\Lambda[g_{\mu\nu}]$ within $O(R^2)$ gravity we have discussed, and they are precisely given by $a-c$ in \eqref{anomaly} with no other way to adjust the parameter.\footnote{The precise proportional factor is beyond our scope because it depends on other terms such as $\Lambda_0$ so they cannot be computed in the power-counting renormalization scheme.} 
Therefore, the quantum renormalization group construction of the bulk action proposed in \cite{Lee:2013xba} is completely consistent with the holographic Weyl anomaly, and it has provided a non-trivial check of the quantum renormalization group origin of the AdS/CFT correspondence.

We have a couple of comments about the agreement.

\begin{itemize}

\item The recipe based on the quantum renormalization group gives only one way to deviate from $a=c$ condition at the curvature squared order from \eqref{anomaly}. In contrast, the bulk action can contain additional two numbers $\lambda_1$ and $\lambda_2$ in \eqref{anomaly}. Presumably, the so-obtained action from the quantum renormalization group is Gauss-Bonnet gravity because it would provide the second order evolution in the radial direction without ghost. Indeed, the holographic renormalization group is best formulated in the quasi-topological gravity, in which the Gauss-Bonnet term is the leading correction, and the inclusion of the other term may require additional care (e.g. how to determine the boundary condition). See \cite{Myers:2010tj} and reference therein for further discussions.

\item The combination $a-c$ is very special in nature. It is the combination that the perturbative string theory can unambiguously compute \cite{Antoniadis:1992sa} and it is related to the holomorphic anomaly in topological string theory. Also, there is a very interesting observation in \cite{Duff:2010ss} about the geometric origin of this particular combination in string compactification.
\item It is possible to generalize the computation with the inclusion of scalars and vectors so that the holographic anomaly for these operators are consistently reproduced from the quantum local renormalization group approach whenever they are not contaminated by the non-universal terms in the power-counting renormalization scheme. The current central charge in $d=4$ and Zamolodchikov metric for the marginal deformations in $d=2$ are good examples.

\end{itemize}

To conclude the article, we present one simple application.
Although it is not directly calculated from the local renormalization group within power-counting renormalization scheme, the consistency of the entire formalism gives a prediction for the metric beta function $\beta_{\mu\nu}$ at $O(R)$ by assuming the theory is dual to the Einstein gravity.\footnote{The metric beta function was first discussed in \cite{Verlinde:1999xm} from the AdS/CFT correspondence.} When $a=c$, the cosmological constant term $\Lambda[g_{\mu\nu}]$ is proportional to $R_{\mu\nu}^2 - \frac{1}{3}R^2$ from the Weyl anomaly \eqref{anomaly}. If the theory is dual to the Einstein gravity without higher derivative terms, these must be cancelled by the term $\beta_{\mu\nu} \beta_{\rho\sigma}G^{\mu\nu;\rho\sigma}$ that appears after integrating out $\pi^{\mu\nu}$. The minimal solution of the cancellation is 
\begin{align}
\beta_{\mu\nu} &= 2g_{\mu\nu} + \sqrt{a}\left(R_{\mu\nu} - \frac{R}{6}g_{\mu\nu}\right)  \cr
{G}^{\mu\nu;\rho\sigma} &=  g^{\mu\rho} g^{\nu\sigma} - g^{\mu\nu} g^{\rho\sigma} \ ,  \label{metricb}
\end{align}
where $G^{\mu\nu;\rho\sigma}$ is again the inverse of the de-Wit metric $\mathcal{G}_{\mu\nu;\rho\sigma} =  (g_{\mu\rho}g_{\nu\sigma} - \frac{1}{3}g_{\mu\nu}g_{\rho\sigma})$ so that $\sqrt{g} \mathcal{G}_{\mu\nu;\rho\sigma} G^{\rho\sigma;\eta\kappa} = \delta_{\mu}^\eta \delta_{\nu}^\kappa$. As a further consistency check, we remark that the metric beta function \eqref{metricb} is a gradient
\begin{align}
\beta_{\mu\nu} = \mathcal{G}_{\mu\nu;\rho\sigma} \frac{\delta S_{\mathrm{EH}}}{\delta g_{\rho\sigma}} \ , \label{gradient}
\end{align}
where $S_{\mathrm{EH}} = \int d^4x \sqrt{g} (12 + \sqrt{a} R)$ is nothing but the Einstein-Hilbert action. This gradient property is necessary to get rid of the first order derivative in $z$ direction \cite{Lee:2013xba} from the bulk action. 

The metric beta function was recently studied from the holographic perspective in \cite{Jackson:2013eqa}\cite{Kiritsis:2014kua} but the relative coefficient may look different from ours. It might be attributed to the scheme choice.\footnote{We have learned that the scheme dependence will be studied in the updated version of \cite{Kiritsis:2014kua} in more detail.}  Our expression agrees with $\dot{g}_{\mu\nu}$ in \cite{Jackson:2013eqa}. One difference from $\beta_{\mu\nu}$ in \cite{Jackson:2013eqa} is that we use the de-Wit metric to lower the indices in \eqref{gradient}. 
In \cite{Kiritsis:2014kua}, the tracelessness condition has been imposed to fix the scaling dimension of the volume element out of $g_{\mu\nu}$. Our principle, instead,
 is the consistency of the quantum local renormalization group and the gradient property, which is naturally imposed in the prescription of \cite{Lee:2013xba}.

The above argument implies that if there were additional $bR^2$ term in the Weyl anomaly \eqref{anomaly}, the cancellation in the higher derivative term would be inconsistent with the gradient property. This in turn means that the bulk theory, if any, would  not  be  invariant under $d+1$ diffeomorphism transformation within pure gravity.
Of course, we know that the $bR^2$ term in the Weyl anomaly does not satisfy the Wess-Zumino consistency condition when the theory is conformal invariant. We also know that when the bulk theory has the full $d+1$ dimensional diffeomorphism, the scale invariant geometry should imply AdS space-time with the full conformal symmetry within pure gravity \cite{Nakayama:2010wx}. In this manner, the whole discussion is mutually consistent. With the inclusion of the additional matter sector, possibly at the sacrifice of unitarity, we may be able to relax the condition, however.
The detailed discussion with the matter will be presented elsewhere.

\section*{Acknowledgements}
I would like to thank S.~S.~ Lee for the discussions and his kind hospitality at Perimeter Institute. I also would like to thank F.~Nitti, E.~Kiritsis and H.~Verlinde for the correspondence. 
This work is supported by the World Premier International Research Center Initiative
(WPI Initiative), MEXT, Japan.

\end{document}